\documentclass[twocolumn,showpacs,preprintnumbers,amsmath,amssymb,APS,prl,nofootinbib,superscriptaddress]{revtex4-1}

\usepackage{hyperref}
\usepackage{dcolumn}
\usepackage{bm}
\usepackage[spanish,english]{babel}
\usepackage{amsfonts}
\usepackage{amssymb}
\usepackage{graphicx}
\usepackage[latin1]{inputenc}

\begin{document}

\title{Nonsingular black holes in quadratic Palatini gravity}

\author{Gonzalo J. Olmo}
\affiliation{Departamento de F\'{i}sica Te\'{o}rica and IFIC, Centro Mixto Universidad de
Valencia - CSIC. Universidad de Valencia, Burjassot-46100, Valencia, Spain}
\author{D. Rubiera-Garcia}
\affiliation{Departamento de F\'{i}sica, Universidad de Oviedo, Avenida Calvo Sotelo 18, 33007, Oviedo, Asturias, Spain}

\begin{abstract}
We find that if general relativity is modified at the Planck scale by a Ricci-squared term, electrically charged
black holes may be nonsingular. These objects concentrate their mass in a microscopic sphere of radius
$r_{core}\approx N_q^{1/2}l_P/3$, where $l_P$ is the Planck length and $N_q$ is the number of electric charges. The singularity is avoided if the mass of the object satisfies the condition $M_0^2\approx m_P^2 \alpha_{em}^{3/2} N_q^3/2$, where $m_P$ is the Planck mass and $\alpha_{em}$ is the fine-structure constant. For astrophysical black holes this amount of charge is so small that their external horizon almost coincides with their Schwarzschild radius. We work within a first-order (Palatini) approach.
\end{abstract}


\date{\today}

\maketitle

\section{Introduction}

The combination of general relativity (GR) and quantum theory leads to important results related with the existence of compact astrophysical objects. When nuclear reactions are no longer effective in the interior of not too massive stars, gravitational collapse can be avoided thanks to the quantum degeneracy pressure of matter, which leads to the formation of white dwarfs and neutron stars \cite{Chandrasekhar-31}. For very massive stars, however, this quantum pressure is unable to counterbalance the gravitational pull and matter collapses yielding a black hole. According to GR, black holes are extremely simple objects that can be fully characterized by their mass, charge, and angular momentum and that are shielded by an event horizon, which generates the emission of quantum particles via Hawking radiation \cite{Hawking,Fabbri:2005mw}. Hidden inside the event horizon, the mass and charge of the collapsed object are concentrated on a point (or a ring if there is angular momentum) of zero volume. In this infinite density point (or ring) curvature scalars diverge, thus implying the existence of a singularity. In such an extreme scenario, one would expect that the quantum properties of the gravitational field could halt the collapse and provide a new stable object in much the same way as white dwarfs and neutron stars arise when the quantum degeneracy pressure of matter dominates. On dimensional grounds, quantum gravitational effects should become non negligible at densities of order $\rho_P\sim c^5/\hbar G^2\approx 10^{94}$ g/cm$^3$ \cite{Shapiro-Teukolsky}. This exceedingly large density sets the scale at which GR should be replaced by some improved description able to stabilize the collapsed object. Due to the lack of a fully developed and tractable quantum theory of gravity, we could gain some insight on how the internal structure of black holes is modified by considering departures from the dynamics of GR at high curvatures. In this sense, extensions of GR formulated in a first-order, or Palatini, approach \cite{Olmo2011a} seem a very promising framework to explore aspects of quantum gravity phenomenology \cite{Olmo2011b,Olmo-Singh}. This type of theories have been studied in cosmological models finding that the big bang singularity can be avoided in very general situations \cite{Bounces}. In these scenarios, the universe began in a contracting phase that bounced off to an expanding phase after reaching a minimum volume characterized by a density of order $\rho_P$. The successful results of that approach suggest that black hole interiors could be modified avoiding the zero-volume singularity of GR.

In this paper we show that electrically charged black holes in an extension of GR defined by the Planck-scale corrected action ($l_P=\sqrt{\hbar G/c^3} \equiv$ Planck length)
\begin{eqnarray}\label{eq:action}
S&=&\hbar \int \frac{d^4x \sqrt{-g}}{16\pi l_P^2}\left[R+l_P^2\left( a R^2+ R_{\mu\nu}R^{\mu\nu}\right)\right]+ \\
&+& S_m[g_{\mu\nu},\psi] \nonumber ,
\end{eqnarray}
formulated \`{a} la Palatini, i.e. by assuming that metric and connection are independent entities \cite{Olmo2011a}, concentrate their mass in a compact sphere of finite volume. Theories of this type are expected to arise when quantum effects are considered in general curved space-times. For external observers, these black holes look essentially like those found in GR, though for a certain charge-to-mass ratio the geometry is nonsingular, which sheds new light on how black hole singularities could be removed by nonperturbative quantum effects. In (\ref{eq:action}) we have denoted $R=g^{\mu\nu}R_{\mu\nu}$, $R_{\mu\nu}=R_{\nu\mu}\equiv{R^\rho}_{\mu\rho\nu}$,  ${R^\alpha}_{\beta\mu\nu}=\partial_\mu\Gamma_{\nu\beta}^\alpha-\partial_\nu\Gamma_{\mu\beta}^\alpha+\Gamma_{\mu\lambda}^\alpha\Gamma_{\nu\beta}^\lambda-\Gamma_{\nu\lambda}^\alpha\Gamma_{\mu\beta}^\lambda$, and  $S_m[g_{\mu\nu},\psi]$ represents the matter action. To remain as close to GR as possible, the independent connection only appears in the gravitational sector of the theory through the definition of ${R^\alpha}_{\beta\mu\nu}$.

\section{Field equations}

Let us now derive the field equations and discuss the peculiar dynamical properties of the theory  (\ref{eq:action}). Taking independent variations of (\ref{eq:action}) with respect to metric and connection yields
\begin{eqnarray}
f_R R_{\mu\nu}-\frac{1}{2}f g_{\mu\nu}+2f_Q R_{\mu\alpha}{R^{\alpha}}_{\nu}&=&\kappa^2 T_{\mu\nu} \label{eq:f(R,Q)-metric}\\
\nabla_\alpha\left[\sqrt{-g}\left(f_R g^{\beta\gamma}+2f_Q R^{\beta\gamma}\right)\right]&=&0 \ , \label{eq:f(R,Q)-connection}
\end{eqnarray}
where $\kappa^2=8\pi l_P^2/ \hbar$, $f=R+l_P^2\left(a R^2+ R_{\mu\nu}R^{\mu\nu}\right)$, $f_R\equiv \partial_R f$, $Q\equiv R_{\mu\nu}R^{\mu\nu}$, and $f_Q=\partial_Q f$. Defining the tensor ${P_\mu}^\nu=R_{\mu\alpha}g^{\alpha\nu}$, one finds that (\ref{eq:f(R,Q)-metric}) establishes an algebraic relation between the components of ${P_\mu}^\nu$ and those of ${T_\mu}^\nu\equiv T_{\mu\alpha}g^{\alpha\nu}$, i.e., ${P_\mu}^\nu={P_\mu}^\nu({T_\alpha}^\beta)$. Once the explicit relation ${P_\mu}^\nu({T_\alpha}^\beta)$ is known, one can express $R$ and $Q$ in terms of the matter according to the identities $R={P_\mu}^\mu$ and $Q={P_\mu}^\alpha{P_\alpha}^\mu$ (see \cite{Bounces,OSAT09} for details and applications), which allows to solve the connection equation (\ref{eq:f(R,Q)-connection}). The independent (symmetric) connection $\Gamma_{\alpha\beta}^\gamma$ turns out to be the Levi-Civita connection of an auxiliary metric $h_{\mu\nu}$ related with the physical metric $g_{\mu\nu}$ as follows
\begin{equation}\label{eq:hmn-general}
{h}_{\mu\nu}=\sqrt{\det \hat{\Sigma}}{\left({\Sigma^{-1}}\right)_\mu}^\alpha g_{\alpha\nu} \ , \ {h}^{\mu\nu}=\frac{g^{\mu\alpha}{\Sigma_\alpha}^\nu}{\sqrt{\det \hat{\Sigma}}}
\end{equation}
where ${\Sigma_\alpha}^\nu=f_R\delta_\alpha^\nu+2f_Q {P_\alpha}^\nu$ is also a function of ${T_\mu}^\nu$ and $\hat{\Sigma}$ is the matrix representation of ${\Sigma_\alpha}^\nu$. It is important to note that when $T_{\mu\nu}=0$ the field equation (\ref{eq:f(R,Q)-metric}) boils down exactly to GR in vacuum. This can be shown by taking the trace of (\ref{eq:f(R,Q)-metric}) with $g^{\mu\nu}$, which for our model gives $R=-\kappa^2 T$ in general and $R_{vac}=0$ in vacuum, and then using the vacuum relation ${P_\mu}^\nu|_{vac}=-\frac{f_R}{4f_Q}\left(1-\sqrt{1+\frac{4f_Q f}{f_R^2}}\right)|_{vac}{\delta_\mu}^\nu$ to obtain $Q_{vac}=0$ and ${P_\mu}^\nu|_{vac}=0$.  From this it follows that in regions where $T_{\mu\nu}=0$ one has ${\Sigma_\alpha}^\nu=\delta_\alpha^\nu$, $h_{\mu\nu}=g_{\mu\nu}$ and $R_{\mu\nu}=0$. This shows that in our theory the dynamics departs from that of GR only in regions that contain sources\footnote{In general, for Palatini $f(R,Q)$ theories the vacuum field equations boil down to GR with an effective cosmological constant.}. Moreover, this fact guarantees that there are no new propagating degrees of freedom and, therefore, the resulting theory is not affected by ghosts or other dynamical instabilities \cite{Olmo2011a}, which clearly manifests the inequivalence between metric and Palatini formulations for higher order gravity (see for instance \cite{Borunda2008} for a discussion on this point, and \cite{Cartan} for a recent attempt to describe Kalb-Ramond fields in terms of torsion in a Palatini-Cartan approach). Thus, in vacuum $h_{\mu\nu}=g_{\mu\nu}$ implies that $\Gamma_{\alpha\beta}^\gamma$ coincides with the Levi-Civita connection of $g_{\mu\nu}$. However, the presence of matter-energy induces a relative deformation between $h_{\mu\nu}$ and $g_{\mu\nu}$, given by (\ref{eq:hmn-general}), such that $\Gamma_{\alpha\beta}^\gamma$ is no longer metric compatible. Physically this means that a given background metric can be deformed locally by the presence of matter in a way that depends on how the energy-momentum density is distributed. This modification of the gravitational dynamics is in sharp contrast with the more standard metric approach, in which the connection is constrained \emph{a priori} to be the Levi-Civita connection of the metric. In that case, the modified dynamics is characterized by the existence of new propagating degrees of freedom which manifest themselves in the form of effective dynamical fields or higher derivatives of the metric. In our theory the modified gravitational dynamics is due entirely to the very presence of matter and its active role in the construction of the connection.

The Schwarzschild black hole is the most general spherically symmetric, non-rotating vacuum solution of GR
and also of (\ref{eq:f(R,Q)-metric}). However, that solution assumes that all the matter is concentrated on a point
of infinite density, which is not consistent with the dynamics of (\ref{eq:f(R,Q)-metric}). In fact, if one considers
the collapsing object as described by a perfect fluid that behaves as radiation during the last stages of the collapse, explicit computation of the scalar $Q=R_{\mu\nu}R^{\mu\nu}$ shows that the energy density $\rho$ is bounded from above by $\rho_{max}=\rho_P/32$, where $\rho_P\equiv 3 c^5/4\pi \hbar G^2$ \cite{OSAT09,Bounces}. Therefore, one should study the complicated process of collapse of a spherical non-rotating object to determine how the Schwarzschild metric is modified in our theory. To avoid such an involved analysis, which would require advanced numerical methods, we consider instead static vacuum space-times with an electric field, which allows to make significant progress using analytical methods only. Disregarding the rotational degrees of freedom, 
static spherically symmetric electrically charged black holes provide the simplest scenarios to address the qualitative and quantitative changes produced by Palatini gravities with Planck-scale corrections on the very internal structure of black holes. The resulting solutions should therefore be seen as Planck-scale modifications of the usual Reissner-Nordstr\"om solution of GR. We thus take
\begin{equation} \label{eq:Maxwell}
S_m[g_{\mu\nu}]= -\frac{1}{16\pi} \int d^4x \sqrt{-g} F_{\alpha\beta}F^{\alpha\beta},
\end{equation}
where $F_{\mu\nu}=\partial_{\mu}A_{\nu} -\partial_{\nu} A_{\mu}$ is the field strength tensor of the electromagnetic potential $A_\mu$. We shall focus on purely electrostatic spherically symmetric configurations, for which the only non vanishing component is
\begin{equation}
F^{tr}=\frac{q}{r^2}\frac{1}{\sqrt{-g_{tt}g_{rr}}} \ ,
\end{equation}
which is a solution of the (sourceless) equations $\nabla_\mu F^{\mu\nu}=0$ assuming the diagonal metric $ds^2=g_{tt}dt^2+g_{rr}dr^2+r^2d\Omega^2$. The associated stress-energy tensor becomes
\begin{equation}\label{eq:Tmn}
{T_\mu}^\nu=\frac{1}{4\pi}\left[{F_{\mu}^{\alpha}} {F}_{\alpha}^{\nu}-\frac{F_{\alpha\beta}F^{\alpha\beta}}{4}\delta_\mu^\nu\right]=\frac{q^2}{8\pi r^4} \left( \begin{array}{ccc}
 -\hat{I} & \hat{0}  \\
\hat{0} & \hat{I}
\end{array} \right) \ ,
\end{equation}
where $\hat{I}$ and $\hat{0}$ are the $2\times 2$ identity and zero matrices, respectively. To solve for the metric, it is convenient to rewrite (\ref{eq:f(R,Q)-metric}) as ${P_\mu}^\alpha {\Sigma_\alpha}^\nu=\frac{f}{2}\delta_\mu^\nu+\kappa^2{T_\mu}^\nu$ and use (\ref{eq:hmn-general}) to get  an equation involving $h_{\mu\nu}$ and the matter only (recall that ${P_\mu}^\nu=R_{\mu\alpha}g^{\alpha\nu}$)
\begin{equation}\label{eq:Rmn-h}
{R_\mu}^\nu({h})=\frac{1}{\sqrt{\det \hat{\Sigma}}}\left(\frac{f}{2}{\delta_\mu}^\nu +\kappa^2 {T_\mu}^\nu\right) \ .
\end{equation}
Inserting (\ref{eq:Tmn}) into (\ref{eq:Rmn-h}) and knowing that in our theory $Q=\frac{(2Gq^2/c^4)^2}{r^8} \equiv \frac{r_q^4}{r^8}$, ${\Sigma_\mu}^\nu={diag}[\sigma_-,\sigma_-,\sigma_+,\sigma_+]$, where $\sigma_\pm\equiv \left(1\pm\frac{l_P^2r_q^2}{r^4}\right)$, we find \cite{OR2011b}
\begin{equation}
{R_\mu}^\nu(h)=\frac{r_q^2}{2r^4}\left(\begin{array}{ccc}
-\frac{1}{\sigma_+} \hat{I}& \hat{0} \\
\hat{0} & \frac{1}{\sigma_-}\hat{I}
\end{array} \right)  \label{eq:Ricci-h4} \ ,
\end{equation}
which recovers GR when $l_P\to 0$.

\section{Metric and Kretchsmann scalar}

Since the right-hand side of (\ref{eq:Ricci-h4}) does not depend on $g_{\mu\nu}$, we can solve directly for $h_{\mu\nu}$ and then use (\ref{eq:hmn-general}) to obtain $g_{\mu\nu}$.
Defining an auxiliary line element for $h_{\mu\nu}$ as $ d\tilde{s}^2=h_{\mu\nu} d\tilde{x}^\mu d\tilde{x}^\nu=-A(\tilde{r})e^{2\Psi(\tilde{r})}dt^2+\frac{1}{A(\tilde{r})}d\tilde{r}^2+\tilde{r}^2d\Omega ^2$
with $A(\tilde{r})=1-2M(\tilde{r})/\tilde{r}$, we find that $\Psi(\tilde{r})=$constant can be absorbed into a redefinition of the time coordinate, like in GR. The metric $g_{\mu\nu}$ can thus be expressed as (with $z\equiv r/\sqrt{r_q l_P}$)
\begin{eqnarray}\label{eq:g}
g_{tt}&=&-\frac{A(z)}{\sigma_+} \ , \ g_{rr}=\frac{\sigma_+}{\sigma_-A(z)}  \\
 A(z)&=&1-\frac{\left[1+\delta_1 G(z)\right]}{\delta_2 z \sigma_-^{1/2}} \nonumber .
\end{eqnarray}
where we used the relation $\tilde{r}^2=r^2 \sigma_{-}$, and defined
\begin{equation}\label{eq:d1d2}
\delta_1=\frac{1}{2r_S}\sqrt{\frac{r_q^3}{l_P}} \ , \
\delta_2= \frac{\sqrt{r_q l_P} }{r_S}
\end{equation}
Here $r_S\equiv 2M_0$, and $M_0$ is an integration constant that coincides with the Schwarzschild mass in the vacuum case. From (\ref{eq:g}) we see that all the information about the geometry is encapsulated in $G(z)$, which satisfies
\begin{equation} \label{eq:Gz}
\frac{dG}{dz}=\frac{z^4+1}{z^4\sqrt{z^4-1}} \ .
\end{equation}
It is easy to see that for $z \gg 1$ we have $G(z)\approx -1/z-3/10z^5$, which leads to $A(r)\approx 1-r_S/r+r_q^2/2r^2-r_S r_q^2l_P^2/2r^5$ and recovers GR when $r\gg l_P$. Note that for the electromagnetic field the function $dG/dz$ is not real for $z<1$. This is a dynamical consequence of the theory, not a coordinate problem, and will be discussed later. In general, (\ref{eq:Gz}) can be exactly solved using hypergeometric functions. Focusing on the small $z$ region, which is the relevant one to look for departures from GR, we find
\begin{equation}\label{eq:G-hyper}
G(z)=\beta +\frac{1}{2} \sqrt{z^4-1} \left[f_{\frac{3}{4}}(z)+f_{\frac{7}{4}}(z)\right] \ ,
\end{equation}
where $f_{\lambda}(z)= {_{2}F}_1[\frac{1}{2},\lambda,\frac{3}{2},1-z^4]$, and $\beta\approx -1.74804$ is a constant resulting from matching the large and small $z$ expansions. Expanding (\ref{eq:G-hyper}) around $z\approx 1$, we find
\begin{eqnarray}
g_{tt}&\approx & \frac{\left(1+\beta  \delta _1\right)}{4\delta _2 \sqrt{z-1}}-\frac{1}{2}\left(1-\frac{\delta _1}{ \delta _2}\right)+O(\sqrt{z-1}) \label{eq:gtt_series} \\
g_{rr}&\approx & -\frac{\delta _2}{\left(1+\beta  \delta _1\right) \sqrt{z-1}}-\frac{2\delta _2\left(\delta _2-\delta _1\right)}{\left(1+\beta  \delta _1\right)^2}-O(\sqrt{z-1}) \nonumber
\end{eqnarray}
The above expansions show that the $g_{tt}$ and $g_{rr}$ components are in general divergent as $z\to 1$. However, they also point out the existence of a particular combination of mass and charge for which the above expansion should be reconsidered. In fact, if we take $\delta_1=-1/\beta\equiv \delta_1^*$, the expansion near $z\approx 1$ becomes
\begin{eqnarray}\label{eq:gtt_constrained}
g_{tt}&\approx &-\frac{1}{2}\left(1-{\delta_1^*}/{\delta _2 }\right)-\left(1-{2\delta_1^*}/{3 \delta _2 }\right)(z-1)+\ldots \\
g_{rr}&\approx & -\frac{1}{2\left(1-{\delta_1^*}/{\delta _2 }\right)}\frac{1}{(z-1)}+\frac{\left(1-7{\delta_1^*}/3{\delta _2 }\right)}{4\left(1-{\delta_1^*}/{\delta _2 }\right)^2}+\ldots  \nonumber
\end{eqnarray}
Though the $g_{rr}$ component in (\ref{eq:gtt_constrained}) diverges as $z\to 1$, the introduction of a new radial coordinate $z^*(z)=2 \sqrt{z-1}/(1-{\delta_1^*}/{\delta _2 })\equiv r^*/\sqrt{r_ql_P}$ turns the physical line element near $z\equiv r/\sqrt{r_q l_P}=1$  into
\begin{equation}
ds^2\approx \frac{1}{2}\left(1-{\delta_1^*}/{\delta _2 }\right)\left[-dt^2+(dr^*)^2\right]+(r_q l_P) d\Omega^2 \ .
\end{equation}
This Minkowskian line element implies that the geometry at $z=1$ is regular. This contrasts with the general case $\delta_1\neq \delta_1^*$ where, defining  $\hat{r}=2\delta_2 r/(1+\beta \delta_1)$, (\ref{eq:gtt_series}) leads to
\begin{equation}
ds^2\approx -\frac{1-\delta_1/\delta_1^*}{4\delta_2\sqrt{z-1}}\left[-dt^2+(d\hat{r})^2\right]+(r_q l_P) d\Omega^2 \ ,
\end{equation}
which is singular at $z=1$. In fact, one finds that the Kretschmann scalar of $g_{\mu\nu}$  blows up at $z=1$ if  $\delta_1\neq \delta_1^*$,
\begin{equation}
{R^\alpha}_{\beta \mu\nu}{R_\alpha}^{\beta \mu\nu}=K_0+(1-\delta_1/\delta_1^*)K_1+(1-\delta_1/\delta_1^*)^2K_2 \ ,
\end{equation}
because the functions $K_1$ and $K_2$ diverge near $z=1$ as $\sim 1/(z-1)^{3/2}$ and $\sim 1/(z-1)^3$, respectively. Since $K_0$ is finite,  near $z=1$ and for $\delta_1=\delta_1^*$ one finds ${R^\alpha}_{\beta \mu\nu}{R_\alpha}^{\beta \mu\nu}\approx \left(16+\frac{88(\delta_1^*)^2}{9 \delta _2^2}-\frac{64\delta_1^*}{3 \delta _2}\right)/r_q^2l_P^2$. The scalars $g^{\mu\nu}R_{\mu\nu}$ and $R^{\mu\nu}R_{\mu\nu}$ of $g_{\mu\nu}$ have a similar behavior \cite{OR2011b}.

\section{Physical properties}

Let us now discuss the physical interpretation of these results. At the surface $z=1$ or, equivalently, $r_{core}=\sqrt{r_q l_P}$, we have found a singularity that can be avoided if and only if the condition $\delta_1=\delta_1^*$ is satisfied. Expressing the charge as $q=N_q e$, where $e$ is the electron charge and $N_q$ the number of charges, we can write $r_q=\sqrt{2\alpha_{em}}N_q l_P$, where $\alpha_{em}$ is the fine structure constant. With this we find that the area of this surface is given by $A_{core}= N_q\sqrt{2\alpha_{em}} A_P$,  where $A_P=4\pi l_P^2$ is Planck's area. This suggests that each charge sourcing the electric field has associated an elementary quantum of area of magnitude $\sqrt{2\alpha_{em}} A_P$. From this it follows that the ratio of the total charge $q$ by the area of this surface gives a universal constant, $\rho_q=q/(4\pi r_{core}^2)=(4\pi\sqrt{2})^{-1}\sqrt{c^7/(\hbar G^2)}$, which up to a factor $\sqrt{2}$ coincides with the Planck surface charge density. Furthermore, the regularity condition $\delta_1=\delta_1^*$ sets the following mass-to-charge relation
\begin{equation}\label{eq:rq-rs}
r_S=\frac{1}{2\delta_1^*}\sqrt{\frac{r_q^3}{l_P}} \ \leftrightarrow  \  \frac{M_0}{(r_ql_P)^{3/2}}=\frac{1}{4\delta_1^*}\frac{m_P}{l_P^3}\nonumber ,
\end{equation}
which indicates that the matter density inside the sphere of radius $r_{core}$ is another universal constant,
$\rho_{core}^*=M_0/V_{core}=\rho_P/4\delta_1^*$,  independent of $q$ and $M_0$.

These results picture a black hole interior fully compatible with the discussion on compact objects presented in the introduction.  Planck-scale physics acts so as to stabilize the collapsed object yielding a compact core of radius $r_{core}$, which contains all the mass in its interior and all the charge on its surface. The latter point is supported by the linear growth of $A_{core}$ with $N_q$  and the fact that the auxiliary geometry $h_{\mu\nu}$ is not defined in the region $r<r_{core}$ ($z<1$). In fact, since the physical geometry, defined by $g_{\mu\nu}$, is completely regular at $z=1$, the impossibility of extending $h_{\mu\nu}$ into $z<1$ indicates a limitation of the electric field to penetrate inside that region. For a description of the geometry in $z<1$ one should  specify the ${T_\mu}^\nu$ of the sources that carry the mass of the core and the charge that generates the external electric field, which would allow to define a new auxiliary metric $\tilde{h}_{\mu\nu}$ able to parameterize the internal geometry of the core (assuming that suitable matching conditions can be found at $z=1$).  Since the regularity condition (\ref{eq:rq-rs}) and cosmological models \cite{Bounces} support that the interior matter density is bounded, we expect the existence of completely regular interior solutions.

From the large $z$ expansion, we saw that the GR solution is a very good approximation for any $r\gg l_P$. This implies that the location of the external horizon of these charged black holes is essentially the same as in GR, i.e.

\begin{eqnarray}
r_+&=&r_S\left(1+\sqrt{1-2r_q^2/r_S^2}\right)/2= \\
&=& r_S\left(1+\sqrt{1-4\delta_1^*/(N_q\sqrt{2\alpha_{em}})}\right)/2,
\end{eqnarray}
where (\ref{eq:rq-rs}) has been used. For a solar mass black hole, Eq.(\ref{eq:rq-rs}) implies that the number of charges needed to avoid the $z=1$ singularity  is just

\begin{equation}
N_\odot=(2r_S \delta_1^* /l_P)^{2/3}/\sqrt{2\alpha_{em}}\approx 2.91 \times 10^{26},
\end{equation}
or $\sim 484$ moles, which is a tiny amount on astrophysical terms. In general, $N_q=N_\odot (M/M_\odot)^{2/3}$ implies that in astrophysical scenarios $r_+\approx r_S$.

\begin{figure}[h]
\includegraphics[width=0.46\textwidth]{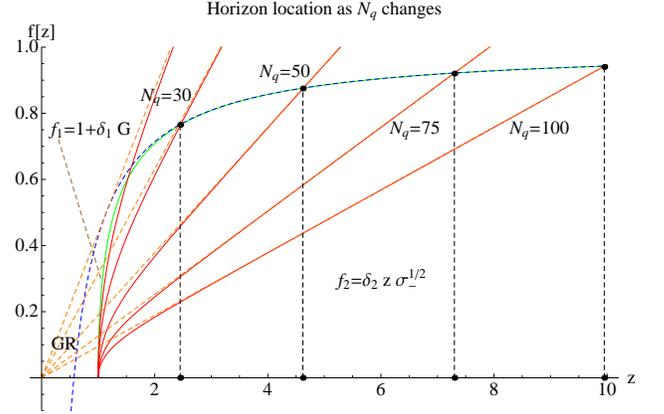}
\caption{The location of the external horizon is given by the intersection of the curve $f_1$ (solid green) with $f_2$ (solid red curves labeled by $N_q$). The dotted curves correspond to GR, $f_1^{GR}$ in blue and the various $f_2^{GR}$ in orange. Color online only.}
\label{fig:1}
\end{figure}

To study how the structure of these nonsingular black holes changes with size, it is useful to express $\delta_2$ as
\begin{equation} \label{eq:regular}
\delta_2=\frac{\delta_1^*}{N_q} \sqrt{\frac{2}{\alpha_{em}}}.
\end{equation}
According to (\ref{eq:g}) the horizons are located at the points where the curves

\begin{equation}
f_1(z)=1+\delta_1^* G(z)
\end{equation}
and

\begin{equation}
f_2=z\delta_1^* \sqrt{2\sigma_-/\alpha_{em}} /N_q
\end{equation}
meet. In GR these curves are just $f_1^{GR}(z)=1-\delta_1^*/z$ and $f_2^{GR}(z)=f_2(z)/\sqrt{\sigma_-}$. In Fig.\ref{fig:1} we see that the cut points for the external horizon are almost indistinguishable from those of GR for $N_q\gtrsim 30$, while the inner horizon is absent. Moreover, from the analysis of these curves, it is found that when $\delta_2>\delta_1^*$, i.e., as $N_q$ drops below the critical value $N_q^c=\sqrt{2/\alpha_{em}}\approx 16.55$ (see Eq.(\ref{eq:regular})), the external horizon disappears and the core becomes directly observable. From (\ref{eq:rq-rs}), the mass of these objects is

\begin{equation}
M=N_q^{3/2} (2\alpha_{em})^{3/4} m_P/(4\delta_1^*)\approx N_q^{3/2}m_P/55,
\end{equation}
and for the particular case $N_q^c$ we find

\begin{equation}
M^c=m_P/(\sqrt{2}\delta_1^*)=\frac{m_P \pi^{3/2}}{3\Gamma[3/4]^2)}\approx 1.23605 m_P.
\end{equation}
This number is also related with the total energy of the electric field in Born-Infeld nonlinear electrodynamics (see e.g. \cite{Gibbons95}). Here it manifests the regularizing role played by gravitation, which forces the electric field to remain in the $z\ge 1$ region. This suggests that the theory (\ref{eq:action}) may provide new insights on the problem of sources in electrodynamics coupled to gravitation \cite{OR2011b}, which already arises in the Reissner-Nordstr\"om and Schwarzschild  solutions of GR \cite{Ortin}. It is worth noting that if $N_q$ is seen as an integer number, besides the core area, the mass of these nonsingular naked cores and black holes is quantized, as has been also recently claimed in \cite{Dvali10a}. One thus expects a discrete spectrum of Hawking radiation, because physically allowed transitions should occur only between regular configurations. This illustrates how Planck-scale physics may affect the perturbative predictions of the semiclassical approach.

\section{Conclusions and perspectives}

We have shown with an exactly solvable model that nonperturbative effects at the Planck scale can generate important modifications on the innermost structure of black holes while keeping their macroscopic features essentially unchanged. The existence of constraints relating the scales of the theory (total mass and charge in our model) to provide nonsingular configurations, and their relation with the quantization of the geometrical structures involved, such as the core surface area, suggests that quantum black holes may have a much richer structure than their classical limits. The impact that our results could have for the experimental search of microscopic black holes  \cite{BHaccelerators} and naked cores makes it necessary a more complete analysis of the influence of the sources on the structure of these compact objects. This and other related studies are currently underway \cite{OR2011b}.

\section*{Acknowledgments}

This work is supported by the Spanish grant FIS2008-06078-C03-02, the Consolider Program CPAN (CSD2007-00042), and the JAE-doc program. Useful comments by A. Fabbri, J. Morales and J. Navarro-Salas are kindly acknowledged.

\end{document}